\documentclass[aps,reprint,a4paper,showpacs,amsmath,amssymb,superscriptaddress,floatfix,footinbib]{revtex4-2}

\usepackage{amssymb}
\usepackage{latexsym}
\usepackage[dvips]{graphicx}
\usepackage{epsfig}

\usepackage{dcolumn}
\usepackage{amsmath}
\usepackage{amsfonts}
\usepackage{bm}
\usepackage{color}

\usepackage[normalem]{ulem}

\usepackage{hyperref}
\hypersetup{colorlinks,linkcolor=blue,urlcolor=blue,citecolor=blue}

\newcommand{\be}{\begin{equation}}
\newcommand{\ee}{\end{equation}}
\newcommand{\bea}{\begin{eqnarray}}
\newcommand{\eea}{\end{eqnarray}}

\newcommand{\p}{\partial}
\newcommand{\s}{\sigma}
\newcommand{\hs}{\hat{\sigma}}

\newcommand{\hS}{\hat{S}}

\renewcommand{\vec}[1]{{\bm #1}}

\begin{document}

\title{Protection of Edge transport in Quantum Spin Hall samples: \\
       Spin-Symmetry Based General Approach and Examples}

\author{Oleg M. Yevtushenko$^*$}
\affiliation{Institut f{\"u}r Theorie der Kondensierten Materie,
             Karlsruhe Institute of Technology, 76128 Karlsruhe, Germany}

\author{Vladimir I. Yudson}
\affiliation{Laboratory for Condensed Matter Physics, National Research University Higher School of Economics, Moscow, 101000, Russia}
\affiliation{Institute of Spectroscopy, Russian Academy of Sciences, Troitsk, Moscow, 108840, Russia}

\date{\today }

\begin{abstract}
Understanding possible mechanisms, which can lead to suppression of
helical edge transport in Quantum Spin Hall (QSH) systems,
attracted huge attention right after the first experiments revealing the fragility
of the ballistic conductance. Despite the very intensive research and the
abundance of theoretical models, the fully consistent explanation of the
experimental results is still lacking. We systematize various theories of
helical transport with the help of the spin conservation analysis which allows
one to single out setups with the ballistic conductance being robustly protected
regardless of the electron backscattering. First, we briefly review different
theories of edge transport in the QSH samples with and without the spin axial
symmetry of the electrons including those theoretical predictions which
are not consistent with the spin conservation analysis and, thus, call for a
deeper study. Next, we illustrate the general approach by a detailed study of
representative examples. One of them addresses the helical edge coupled to
an array of Heisenberg-interacting magnetic impurities (MIs) and demonstrates
that the conductance remains ballistic even if the time-reversal symmetry on the
edge is (locally) broken but the total spin is conserved.
Another example focuses on the effects of the space-fluctuating spin-orbit
interaction on the QSH edge. It reveals weakness of the protection in several
cases, including, e.g., the presence of either the U(1)-symmetric, though not fully
isotropic, MIs or generic electron-electron interactions.
\end{abstract}

\date{\today}


\maketitle

\section{Introduction}

The theory predicting existence of time-reversal invariant two-dimensional (2D)
topological insulators \cite{HasanKane,QiZhang,TI-Shen}, often dubbed Quantum
Spin Hall (QSH) samples \cite{KaneMeleZ2,KaneMeleQSH,bernevig_QSH_2006},
provoked a huge interest because of its physical beauty and high potential for
applications in nanoelectronics, spintronics, and quantum computers. It was realized
\cite{BHZ_2006,WuBernevigZhang} that the one-dimensional (1D) boundary between
trivial and topological insulators, which are distinguished by inversion
of the gap in the spectrum of electrons, can host the so-called helical
gapless edge modes.
Modes of the given helicity possess lock-in relation
between their spin and momentum.
The Kramers degeneracy guarantees that there are always two
counterpropagating modes possessing the same energy and opposite spins.
A single-particle scattering, which converts the
helical mode into its counterpart from the Kramers doublet, is not allowed
without a spin-flip.
Thus, at least in the absence of interactions, the helical modes are
not liable to backscattering and localization caused by material imperfections.
This could provide a possibility to sustain
ballistic transport, which possesses the quantized ballistic conductance,
$ G_0 = e^2 / h $ per one QSH edge, in long 1D conductors. The r{\'o}le of
the topologically nontrivial bulk and of time reversal symmetry (TRS) is often
reflected in referring to the QSH edge transport as ``topologically protected
transport'' or ``transport protected by TRS''.

Transport properties of real samples appeared to be much
more complicated than the above described single-particle picture.
It
became clear already after first measurements that helical QSH transport remains
ballistic only in relatively short samples and the conductance may be smaller
than $ G_0 $ if the edge length is larger than some critical value
\cite{Molenkamp-2007,konig_2013,EdgeTransport-Exp1,EdgeTransport-Exp2}. This points
out to the presence of not-single-particle backscattering of the helical
fermions, which can occur without changing the topological state
of the bulk or breaking TRS. The subballistic conductance demonstrates
incompleteness of the notion of topologically/TRS protected transport
for the interacting helical edges.

The elegancy of the physics of the 2D topological insulators together with
promising potential of their application gave a strong push to
numerous experimental
\cite{roth_2009,EdgeTransport-Exp0,gusev_2011,brune_2012,gusev_2013,nowack_2013,suzuki_2013,gusev_2014,ma_2015,olshanetsky_2015,tang_2017,fei_2017,li_2017,bendias_2018,wu_2018,Piatrusha-TI-TRS,stuhler_2020}
and theoretical
\cite{XuMoore,MaciejkoLattice,CheiGlaz,vayrynen_2016,vayrynen_2017,wang_2017,Klinovaja_Loss_2017,Feiguin_Martins_2017,RKKY-HLL-Bulk,Yevt_2018,hsu_2018,cangemi_2019,TI-KI-noise,hsu_2021}
studies, see also references to other papers which are discussed below in more detail.
Despite enormous efforts of researchers, the fully consistent
theory explaining suppression of QSH helical transport in all
experimental setups is still absent and
remains the hot topic. On the other hand, an abundance
of various theoretical models and some confusions in the terminology
addressing the origin of protected transport
hamper
comprehension of the field.

We would like to somehow brighten such a picture. In this Paper,
we discuss helical transport in the QSH samples with- and without
the axial spin symmetry of the electrons. We emphasize that, if this
symmetry is present, the (sub)ballistic nature of the conductance
is fully determined by the (non)conservation of the total spin projection
on the axis of the spin symmetry.
Importantly, if the total spin projection is conserved, the helical conductance
is ballistic regardless of the electron backscattering and of the presence of
the TRS on the QSH edge. If, on the contrary, this projection is not conserved,
the helical conductance can be suppressed even if the TRS is not broken.

The spin conservation analysis
in systems
with broken spin axial
symmetry is generically more sophisticated. Nevertheless,
we will show that such an appro\-ach remains a useful tool.
In particular, the spin axial symmetry is effectively restored in the
experimentally relevant low-temperature limit, and the analysis of the
spin conservation allows one to identify the setups where the
ballistic conductance can(not) be suppressed.

The paper is organized as follows: the Landauer setup for the helical conductors
and basic notations are introduced in Sect.II. Sect.III and IV are devoted to the
discussion of the QSH systems with- and without the spin axial symmetry of
the electrons, respectively. Sect.III starts with a methodological explanation
of the connection between the spin relaxation and the suppression of the helical
conductance. Both Sect.III and IV include a brief (though not an exhaustive) review of
representative models and theories, which illustrates this connection, and a detailed study
of a corresponding example (with- and without the spin axial symmetry), which
corroborates the above generic statements and the approach. In the review-like
subsections, we critically analyze some contradictive claims on a suppression
of the ballistic conductance in the spin preserving setups. This demonstrates
that using the spin conservation analysis might help to avoid such inconsistencies.

The example of Sect.III addresses scattering of the helical electrons
caused by either uncorrelated or Heisenberg-interacting magnetic impurities (MIs).
They are able to cause backscattering of the individual electrons which is
accompanied by the spin flip. We argue that such a backscattering has no influence
on the linear dc helical conductance if the total spin is conserved. This holds
true regardless of a spin ordering which breaks the TRS on the helical edge.

In the example of Sect.IV, we analyze the protection of helical transport against
combined effects of the space-fluctuating spin-orbit interaction (SOI) and the scattering
of the interacting helical electrons by the MIs. A fully isotopic and short-range
correlated array of the MIs is unable to suppress the helical conductance even
if the SOI fluctuates. On the other hand, the absence of the
spin axial symmetry of the electrons leads to the suppression of the helical
conductance by the U(1)-symmetric, though not fully isotropic, MIs. This follows from
the spin conservation analysis which is applicable despite the absence of the spin
conservation in the laboratory frame.

We conclude the paper by a brief Summary.

\section{Landauer setup for a helical wire}

\begin{figure}[t]
\begin{center}
   \includegraphics[width=0.48 \textwidth]{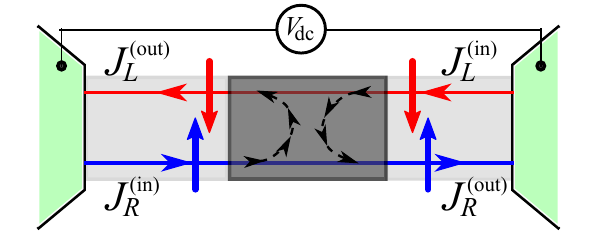}
\end{center}
\vspace{-0.5 cm}
   \caption{
        \label{BlaclBoxSetUp}
        A generic two terminal setup for measuring the helical Landauer conductance:
        1D wires (light grey boxes) with two modes of the same helicity (red and blue lines)
        are connected to 2D leads (colored trapezia). For simplicity, we have
        illustrated the case where the electron spin quantization axis is well-defined
        in the wires. The central part -- the scattering region -- is a conducting
        region where backscattering (black dashed lines) of the electrons
        is possible. The dc voltage, $ V_{\rm dc} $, is applied to the 2D leads.
           }
\end{figure}

We focus on the helical modes of the single QSH edge and discuss the linear dc conductance
of the system. We conventionally subdivide the edge into three parts: two ideal 1D helical
wires connecting leads and a complex scattering region (SR), see Fig.\ref{BlaclBoxSetUp}.

Left and right wires are ideal, i.e. contain no sources for the scattering of the helical
electrons. The middle part, the SR, is a conducting region where the helical electrons can
be scattered.
This corresponds to the standard Landauer setup: The SR acts as a composite scatterer
while left and right ideal helical wires are identical and play a role of contacts connecting
the scatterer and external 2D leads.

\subsection{ Backscattering inside the SR}

The necessary, though not sufficient, condition for suppression of the ballistic conductance is
backscattering of the helical electrons inside the SR.
Since the single particle backscattering of the helical
electrons by a potential disorder is not allowed, it can be caused only by interactions.
The nature of these interactions is broad and ranges from various exchange interactions, to
electron-electron interactions, to a nontrivial (e.g. fluctuating in space)
SOI, to name just a few. Exchange interactions may include those with frozen or dynamical
localized spins, the MIs. The latter are often referred to as Kondo impurities 
even in the weak coupling regime where the Kondo physics is unimportant.

We emphasize that, below, we discuss many examples where backscattering of individual
electrons does not suppress the ballistic conductance since transport remains protected.
This paradoxical situation is the well-known property of the helical systems. It compels
us to distinguish notions of ``backscattering'' and ''the ballistic transport'', see also
the discussion in Sect.\ref{TranspVsBackscatt}.

The analysis of the helical conductance in the QSH sample depends on the presence or absence
of the electron spin axial symmetry. These two cases are discussed in next Sections.

\section{QSH systems with \\
         spin axial symmetry of electrons}

The focus of this section is on the QSH systems with the spin axial symmetry of the electrons.
Namely, we discuss the systems where the spin projection on a given direction is conserved
for all non-interacting helical electrons.
The Hamiltonian describing free helical 1D Dirac fermions in the left/right wires, Fig.\ref{BlaclBoxSetUp},
reads as:
\be
\label{H0}
       \hat{H}_0 = - i v_F \int {\rm d} x \ \hat{\Psi}^{\dagger}(x) \hat{\s}_z \partial_x \hat{\Psi}(x) \, .
\ee
Here $ \, v_F $ is the Fermi velocity; $ \hat{\Psi} = \{ \psi_\uparrow, \psi_\downarrow \}^{\rm T} $ is the spinor
constructed from the fermionic fields with given spin projections on z-axis; the Pauli matrix $ \, \hat{\s}_{z} \, $
(and $ \, \hat{\s}_{x,y} \, $ introduced below) operates in the spinor space.
Eq.(\ref{H0}) is the simplest model which describes the edge modes, for example, in those HgTe/CdTe quantum-well
heterostructures which possess the axial and inversion symmetry around the growth axis.

The Hamiltonian (\ref{H0}) of the free electrons conserves z-component of their spin which is the starting
point for the analysis of the helical conductance. We emphasize that, in the QSH systems with the electron
spin axial symmetry, the spin projection unambiguously defines the chirality of right ($ R $) and left ($ L $)
moving electrons with a given helicity. In this case, one can use subscripts $ \uparrow,\downarrow $ and
$ R,L $ on equal footing.


\subsection{Spin (im)balance}


\subsubsection{(Im)balance of currents}

The Landauer setup shown in Fig.\ref{BlaclBoxSetUp} involves incoming, $ J_{R,L}^{\rm (in)} $, and
outgoing, $ J_{R,L}^{\rm (out)} $, chiral charge currents. Assuming reflectionless contacts between the wires and the
leads, the incoming currents are fully governed by the leads: $ J_{R,L}^{\rm (in)} = e v_F \sum_k n_{R,L}(k)$, with
$ e $ and $ k $ being the electron charge and momentum. The electronic distribution functions $ n_{R,L} $
are determined by the left and right leads, respectively. Outgoing currents are a priori not known.

We consider only those SRs where the charge is conserved and the following charge current balance always
holds true:
\be
\label{Charge-U1}
   J_R^{\rm (in)} + J_L^{\rm (in)} = J_R^{\rm (out)} + J_L^{\rm (out)} .
\ee

Similar to the charge currents, we introduce incoming, $ S_{R/L}^{\rm (in)} $, and outgoing,
$ S_{R/L}^{\rm (out)} $, currents of the electron spin projected on the quantization axis. The interaction
induced backscattering can (or cannot) violate the spin conservation in the SR. Therefore, we cannot
presume a robust balance of the spin currents. Possible violation of the spin conservation can be
characterized by the spin imbalance: $ \delta S = ( S_{R}^{\rm (in)} + S_{L}^{\rm (in)} ) -
( S_{R}^{\rm (out)} + S_{L}^{\rm (out)} ) $. Using the well-known relation between the charge
and spin currents in the QSH systems with the axial symmetry and the certain helicity on
the edges, $ S_{R}^{\rm (in/out)} = J_{R}^{\rm (in/out)} / 2 e $, and $ S_{L}^{\rm (in/out)} =
- J_{L}^{\rm (in/out)} / 2 e $, we express $ \delta S $ via $ J_{R/L}^{\rm (in/out)} $:
\be
\label{Spin-U1}
   \delta S = \left( J_{R}^{\rm (in)} - J_{L}^{\rm (in)} -
                     J_{R}^{\rm (out)} + J_{L}^{\rm (out)} \right) / 2 e.
\ee
It follows form Eq.(\ref{Spin-U1}) that, if the SR is spin conserving, the helical conductance remains ballistic.
Namely, if $ \delta S = 0 $, then [recalling Eq.(\ref{Charge-U1})]
we obtain:
\be
\label{IdealTransp}
   J_\mu^{\rm (in)} = J_\mu^{\rm (out)} , \ \mu = R, L .
\ee
Therefore, the total electric current through the system,
$ J_{\rm tot} = J_R^{\rm (in)} - J_L^{\rm (out)} = J_R^{\rm (out)} - J_L^{\rm (in)} $,
can be expressed via incoming currents:
\be
   J_{\rm tot} = J_R^{\rm (in)} - J_L^{\rm (in)} ,
\ee
which are fully determined by the external voltage $ V_{\rm dc} $ and by contacts between
the external leads and the clean helical wires. Thus, if $ \delta S = 0 $,
the Landauer conductance, $ G_{\rm dc} = J_{\rm tot} / V_{\rm dc} $, is not sensitive to the presence
of the SR in the circuit. In the case of the reflectionless contacts \cite{Datta}, one arrives at
$ G_{\rm dc} = G_0 $ regardless of internal details of the spin (and charge) conserving
SR, including the interaction induced backscattering inside the SR.

If the SR is not spin-conserving and there is the spin imbalance, $ \delta S \ne 0 $, the backscattering
in the SR results in the so-called backscatering current, $ J^{\rm (BS)}_{R,L} = J_{R,L}^{\rm (out)} -
J_{R,L}^{\rm (in)} $, which suppresses the ballistic helical conductance.

\subsubsection{Ballistic transport vs. backscattering \label{TranspVsBackscatt}}

We come across an unusual situation where
the helical Landauer conductance is ballistic despite the interaction induced backscattering
inside the spin preserving SR. Therefore, we should distinguish concepts of ballistic transport
and backscattering and refer to the regime $ G_{\rm dc}= G_0 $ as ``ballistic'', though the helical
electrons can be backscattered. This remarkable robustness of the ballistic conductance with respect
to backscattering is the direct consequence of helicity and
makes the physics of the helical 1D wires distinct from that of usual (non-helical) 1D conductors
\cite{Giamarchi} where any
backscattering efficiently suppresses dc transport.

\subsubsection{Conditions for the spin conservation}

The explicit calculation of the currents is often a very complex task. A feasible and physically
transparent alternative is provided by the analysis of the spin conservation in the SR.

An obvious sufficient condition for the spin conservation is the global spin U(1) symmetry,
$ \exp(i \hat{S}_z \alpha) \hat{H}_{\rm tot} \exp(-i \hat{S}^{\rm tot}_z \alpha) = \hat{H}_{\rm tot} $,
with $ \hat{S}^{\rm tot}_z $ and $ \hat{H}_{\rm tot} $ being
the total (of the SR and of the wires) spin operator and Hamiltonian respectively.
This requirement is sufficient but, generally speaking,
not necessary, cf. Sect.\ref{FluctSOI}.

Several specific examples of spin preserving and non-preserving setups are given in the next
Subsection. Below, we use terms ``the spin U(1) symmetry'' and the above discussed ``spin conservation''
on equal footing.

\subsection{Spin (non)preserving setups \protect\\
            and helical transport}


{\it Interaction with a single magnetic impurity and nanomagnets}:
An exchange interaction of the helical electrons with the MI can result in
backscattering accompanied by the spin-flip. That is why the MIs were considered
as a serious obstacle for ballistic helical transport. The study of the Kondo
physics for the XXZ-anisotropic MI immersed in
the helical Luttinger liquid erroneously predicted suppression of the helical
conductance at finite temperatures \cite{MaciejkoOregZhang}. In reality, such
an impurity backscatters the helical electrons but does not break the spin U(1)
symmetry and, therefore, cannot influence the dc conductance \cite{FurusakiMatveev}.
XYZ-anisotropic MIs violate the spin conservation and are able to suppress the ballistic
conductance but only if they are not Kondo screened. This requires either
the temperature being larger than the Kondo temperature, $ T > T_K $,
or a large value of the Kondo spin, $ S > 1/2 $
\cite{kurilovich_2017,TI-Anisotr-KI}.
The Kondo screening of the XYZ-anisotropic spin-1/2 MI restores the spin conservation at $ T \to 0 $ and,
thus, neutralizes the destructive effect of such an impurity on the ballistic conductance
\cite{vinkler_2020}: $ G_{\rm dc} $ becomes equal to $ G_0 $ at $ T \ll T_K $.

Coupling of the MI to additional degrees of freedom
can result in a finite spin relaxation rate of the MI. It violates the applicability of the
spin conservation analysis and might lead to the subballistic conductance. Some examples
of this effect have been reported in Refs.\cite{Klinovaja_Loss_2017,hsu_2018}:
the hyperfine interaction couples the helical electrons to the nuclei spins located
near the edge. The latter are, in turn, coupled to the spins in the bulk. This
provides a ``leakage'' of the spins to the bulk and suppresses the helical conductance at very
low temperatures \cite{Klinovaja_Loss_2017}. Related effects are substantially material-dependent
and are often very weak. For instance, in such an archetypal material as HgTe/(Hg,Cd)Te, they
are expected to be pronounced only in long samples, $ L \gtrsim 4 \mbox{mm} $, at ultralow
temperatures, $ T \lesssim 5 \mbox{mK} $ \cite{hsu_2018}. Since we address more universal
mechanisms, we follow the majority of studies in the field and do not take into account additional
(beyond those caused by the electron-spin exchange coupling) spin relaxations in the present paper.

The consideration of the MI, which backscatters the helical electrons
in the SR, can be easily extended to macroscopic magnetic scatterers consisting
of a finite ensemble of dynamically interacting spins. For example,
instead of the single MI, one can consider a macroscopic
nanomagnet which can be phenomenologically described by a time-dependent
vector of magnetisation \cite{meng_2014}. If there is no spin relaxation
inside the magnet, the conductance reaches $ G_0 $ after some transient
governed by accumulation of an excessive spin, which flows into the magnet
from the biased helical wires.

An attempt to develop a microscopic theory of the nanomagnet attached
to the helical edge has been presented in Ref.\cite{novelli_2019}. The
authors of this paper have used the Kane-Mele-Hubbard model, where
the nanomagnet is expected to appear spontaneously at small temperatures
near a dislocation placed close by the helical edge. The microscopic
origin of this effect is the electron interaction. By employing
a self-consistent mean-field approximation (MFA), one can come to a doubtful
conclusion on a suppression of the helical conductance by such a nanomaget
even at $ T \to 0 $ \cite{novelli_2019}. Indeed, since the magnetization
is frozen (the basic assumption of the MFA) it certainly violates the spin
conservation and can suppress the helical conductance. However, the Hamiltonian
of the Kane-Mele-Hubbard model has the global spin U(1) symmetry and, therefore,
the spin conservation analysis
puts the MFA prediction in question.
We believe that
fluctuations are crucially important since they bring to life dynamics of
the effective nanomagnet which could restore the ballistic value of the
conductance. Thus, this interesting model deserves a further study.

A detailed and more rigorous microscopic study of the influence of the
MI array (correlated or uncorrelated) on the helical conductance
is presented in the next Subsection. We will show that our microscopically
obtained results are in perfect agreement with the spin conservation analysis.

\vspace{0.25 cm}

{\it Interaction between electrons and two-particle backscattering}: Various electron
interactions can yield two-particle backscattering which is able to suppress
the ballistic conductance if the spin conservation is violated. For example,
anisotropic electron spin interactions on a lattice can govern the umklapp two-electron
backscattering which suppresses helical transport \cite{WuBernevigZhang}.
The electron-electron interaction in quantum dots (charge puddles) attached
to the helical edge was considered in Refs.\cite{vayrynen_2013,vayrynen_2014}.
The spin non-conserving interaction of the electrons inside the dots provides the
spin relaxation and makes the helical edge conductance subballistic.

\vspace{0.25 cm}

{\it Contacts between the helical wires and the SR}:
One should keep in mind that the contacts between the helical wires and the SR
themselves can violate conservation laws and suppress ballistic transport.
For instance, one could erroneously surmise that
the SR containing usual (non-helical) 1D spinful modes and a spinless
disorder does not violate the spin conservation but, nevertheless, suppresses the
conductance via the mechanism of Anderson localization. As a matter of fact, there is no
contradiction since any connection of helical- and non-helical wires in a series
violates validity of the conservation laws. Namely: (i) The contact,
where the spin is preserved, would violate unitarity of the S-matrix (i.e.,
would violate the charge conservation) describing such a connection. This directly
follows from a different number of channels with a given spin in the non-helical- and
helical wire, e.g., $ R_\uparrow $ and $ L_\uparrow $ in the non-helical wire and
only $ R_\uparrow $ in the helical one. (ii) The only other possibility for the
connection in series is to allow the helical electron to change its spin inside the
contact. Clearly, this possibility
is
incompatible with the total spin conservation.

\subsection{Helical modes interacting with a Kondo-Heisen\-berg array - a case study
           \label{e-KHA}
           }

Let us now elaborate at the microscopic level how the above explained spin
conservation analysis
works in the case of the helical edge coupled to an array of the MIs.
Such systems regain at present an increased attention since they can
provide a platform for the realization of chiral Majorana fermions \cite{shamim_2021}.
The MI can be correlated either via the indirect exchange interaction
supported by the itinerant electrons [usually referred to as "the Kondo
array" (KA)] or via the both indirect and direct Heisenberg interaction [usually
named "the Kondo-Heisenberg array" (KHA)].

The influence of the dense KA on the conductivity of the helical edge
was studied in Refs.\cite{AAY,YeWYuA}. The high density
of the MIs means that the Kondo effect is overwhelmed by the MI correlations \cite{doniach_1977}
and the MIs remain unscreened at low temperatures. It was shown that if the helical electrons are
scattered by the KA with the random XYZ-anisotropy one comes across Anderson localization
of the charge carriers \cite{AAY} while, in the spin U(1)-symmetric setup, the backscattering
leads only to the renormalization of the Drude peak in the dc conductivity.
Based on a straightforward though a kind of superficial analogy with the physics of usual
(not helical) interacting 1D wires \cite{MaslovStone,Ponomarenko,SafiSchulz}, a ballistic nature
of the dc conductance in the spin preserving setup was conjectured in Ref.\cite{AAY}. We now
demonstrate that this conjecture is correct and extend the theory to the dense KHA where
the Heisenberg interaction of the MIs may govern the TRS-breaking spin ordering.

\begin{figure}[t]
\includegraphics[width=0.475 \textwidth]{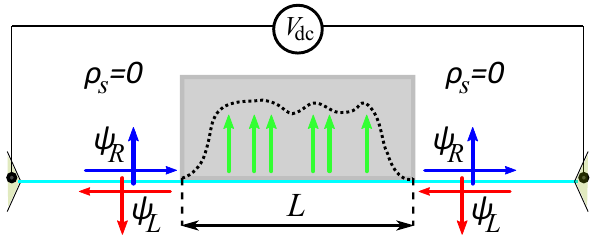}
   \caption{\label{KHLSetUp} (color on-line)
        1D wire with helical electrons, $ \psi_{R/L} $ or equivalently $ \psi_{\uparrow/\downarrow} $,
        coupled to the Kondo-Heisenberg array (green arrows)
        located inside the SR (shadowed rectangle). Density of the MIs (dotted line)
        vanishes outside the SR and fluctuates in space inside it. The interaction constants,
        $ J_{K,H} $ can also depend on the coordinate.
                 }
\end{figure}

The total Hamiltonian describing the helical edge modes, which interact with the KHA
(see Fig.\ref{KHLSetUp}), is $ \hat{H}_{\rm full} = \hat{H}_0 + \hat{H}_H + \hat{H}_{b} +
\hat{H}_{V} $. It includes the fermionic and MI parts, $ \hat{H}_0 $ and $ \hat{H}_H $,
respectively, and the voltage source part $ \hat{H}_{V} $ which depends only on the fermionic
operators. $ \hat{H}_{b} $ is the Hamiltonian of the fermion-MI interaction which describes
backscattering induced by MIs. In the spin-conserved setup, it reads as follows:
\begin{equation}
\label{Hb}
      \hat{H}_{b} = \sum_j \!\! \hat{S}^+_j [J_{K} \psi^{\dagger}_{\downarrow} \psi_{\uparrow}](x_j) + h.c.
\end{equation}
Here
the sum runs over MI positions $ x_j $, $ \, J_{K} \, $
is the position dependent coupling constant between the MIs and the fermions;
$ \, \hS^{\pm}_j = e^{\pm 2 i k_F x_j} ( \hS_x \pm i \hS_y )_j \, $ are rotated  raising/lowering operators of the
MIs. Eq.(\ref{Hb}) corresponds to the U(1)-symmetric XXZ-coupling with $ J_{K} = J_K^{(x)} = J_K^{(y)} $. To simplify
discussion of this example, we take into account neither electron-electron interactions nor
forward-scattering generated by the component $ J_K^{(z)} $  though including them is straightforward and does
not change our conclusions.

The spin Hamiltonian $ \hat H_H $ describes the direct Heisenberg exchange interaction (of quantum
Ising type) between the z-components of the MI spins:
\begin{equation}
\label{Hh}
  \hat H_H =  \sum_j J_H(x_j) \hS_z(j) \hS_z(j+1) \, ;
\end{equation}
where $ J_H $ is the position dependent coupling constant.
For concreteness, we have chosen the exchange interaction between nearest neighbors.
The inter-impurity distance $ \xi_s(j) = x_{j+1} - x_j $ can fluctuate in space
if the spin array is geometrically disordered.

The MI density is $ \rho^{(d)}_s(x) = \sum_j \delta(x - x_j) $ for the discrete array. Its smeared
counterpart, $ \rho_s(x) $, vanishes outside the SR and is finite and coordinate-dependent
inside it.

Our starting point is similar to that of Refs.\cite{MaslovStone,MaslovDurtyLL}: We express
the electric current via a convolution of the non-local conductivity, $ \sigma(x,x';\omega) $, and an
inhomogeneous electric field $ \, E(\omega,x') $:
\begin{equation}
   j(x,t) = \int {\rm d}x' \int  \frac{{\rm d} \omega}{2 \pi} e^{- i \omega t} \sigma(x,x';\omega) E(\omega,x') \, .
\end{equation}
$ E(\omega,x) $ is governed by the applied voltage.
Next, we bosonize the theory, use the technique of the functional integrals
on the imaginary time, and describe the fermions by the standard Lagrangian of the helical
Luttinger Liquid with the source term \cite{Giamarchi,WuBernevigZhang}:
\be
\label{L_HLL}
  {\cal L}_{\rm HLL} = \left[ (\partial_\tau \phi)^2 +  (v_F \partial_x \phi)^2 \right] / 2 \pi v_F + i \chi \phi \, .
\ee
The Fourier transform of the bosonic Green's
function (GF), $ G(x,x'; \tau) = - \langle \phi(x',\tau) \phi(x,0) \rangle $, yields the Matsubara conductivity:
$
   \sigma(x,x'; \bar{\omega} ) = ( e^2 \bar{\omega} / \pi^2 ) \, G(x,x'; \bar{\omega} ) \, .
$
The Lagrangian describing the spin-conserving backscattering reads:
\begin{equation}
\label{L_bscatt}
   {\cal L}_{b}[n_z,  \alpha, \phi] = 2 s \rho_s(x) J_{K}(x) \sqrt{1 - n^2_z} \cos( \alpha - 2 \phi ) \, .
\end{equation}
Here we parameterize each localized spin by its azimuthal
angle, $ \, \alpha $, and projection on $z$-axis, $ \, |n_z| \le 1 $:
$
    S^{\pm} = s e^{\pm i \alpha} \sqrt{1 - n^2_z} \, , \ S_z = s n_z;
$
with
$ s $ being the spin value. This parametrization requires the Wess-Zumino term in the Lagrangian
\cite{NagaosaBook,ATsBook}:
\be
  {\cal L}_{\rm WZ}[ n_z,  \alpha ] = i s \rho_s (1 - n_z ) \partial_\tau \alpha \, .
\ee


Let us now shift the spin phase $ \tilde{ \alpha} = \alpha - 2 \phi $. The full Lagrangian
in the new variables is
\begin{equation}
\label{Lagr}
  {\cal L} = {\cal L}_{\rm HLL}[\phi, \chi] + {\cal L}_{\rm WZ} [n_z, 2 \phi] + {\cal L}_{\rm MI}  [ n_z, \tilde{\alpha} ] \, .
\end{equation}
Here $ {\cal L}_{\rm MI} \equiv {\cal L}_{b}[n_z, \tilde{ \alpha}, 0] + {\cal L}_{H} [n_z] + {\cal L}_{\rm WZ} [n_z, \tilde{ \alpha}] $
is the MI Lagrangian; $ {\cal L}_{H} $ describes the Heisenberg interaction of the MIs.

The coupling between the fermionic and MI sectors is reflected in Eq.(\ref{Lagr}) only
by $ {\cal L}_{\rm WZ} [n_z, 2 \phi] $. Its contribution
vanishes in the dc limit
and the MI variables drop out from the equation for the dc conductivity which reduces
to that of the clean helical wire, see the proof in \appendixname~A.
Therefore, coupling between the helical modes and the spin-preserving KHA is unable
to change the ballistic dc helical conductance as one expects from the above explained generic arguments.


Dense and large KHAs can host various spin orders \cite{KH-CSL}.
If $ J_H $ exceeds a critical value, $ J_H^* $,
the crossover (the quantum phase transition in the infinite system)
of the Ising type occurs at zero temperature and the component $ S_z $
on the edge of the QSH sample acquires a non-zero semiclassical average value,
$ s \langle n_z \rangle $ \cite{TI-SupSol}.
This means that
the TRS can be spontaneously broken on the QSH edge
\footnote{We mean the experimentally relevant large time scale
while discussing the finite Kondo-Heisenberg array inside the SR.}.
Nevertheless, the helical dc transport remains ballistic and protected
as soon as the total spin is conserved. These arguments
show the secondary importance of the TRS on the edge for the protection of the helical transport.
Note that the fermion backscattering by the magnetically ordered KHA
requires an energy of the order of the domain wall energy, $ {\cal E}_{\rm DW} $
\footnote{The domain wall energy can be estimated as $ {\cal E}_{\rm DW} \sim
[(J_H - J_H^*) / J_K]^2 \xi_s / v_F $ close to the critical region, see details in
Ref. \cite{TI-SupSol}, and $ {\cal E}_{\rm DW} \sim J_H $ at $ J_H \gg J_K $.}.
Therefore, it is expected to be suppressed in regimes where all characteristic energy scales
are $ \ll {\cal E}_{\rm DW} $.

We emphasize that the dc helical transport in the current example is
sensitive neither to the profile of the spin density, $ \rho_s(x) $, nor to the spatial
inhomogeneity of the coupling constants, $ J_{K,H}(x) $.
This is the exclusive property of the helical system. On the other hand,
the ac conductance might be sensitive to details of the setup.

Let us conclude this Section with an example which demonstrates a predictive power
and usefulness of the spin conservation analysis. So far, we have taken into account
only the Heisenberg interaction between z-components of the MI spins. The presence of
the spin-preserving in-plane Heisenberg interaction of the MIs,
$ {\cal L}_H^{\rm (xy)} \propto \rho_s e^{- 2 i k_F \xi_s} S^+(x + \xi_s) S^-(x) + h.c. $,
makes the analysis of the dc conductance technically very complicated. In particular,
the phase $ \phi $ cannot be gauged out simultaneously from the Lagrangians $ {\cal L}_b $
and $ {\cal L}_H^{\rm (xy)} $. Qualitative arguments confirming the helical ballistic
conductivity are given in Ref.\cite{TI-SupSol} for the case of an infinite system where
the total spin is conserved. However, based on the spin conservation analysis, one can
state without explicit calculations that the conductance remains ballistic in this complicated
setups because the total spin is conserved.

\section{QSH systems with broken \\
         spin axial symmetry of electrons}

Another class of the QSH systems embraces those TRS-invariant 2D topological insulators where
the spin axial symmetry of the electrons is broken. As a result, the electron spin quantization
axis cannot be uniquely defined for all helical electrons and the z-component of the electron
spin is no longer conserved.
Well-known microscopic mechanisms, which are compatible with the TRS and break the spin
axial symmetry, are the bulk- and the structural inversion asymmetry \cite{liu_2008,rothe_2010}.
The effects of the broken axial symmetry can be described by
an effective spatially homogeneous Rashba SOI in the 2D bulk. Close to the 1D edge of the QSH
sample, the Rashba SOI can be inhomogeneous and strongly fluctuate.

The absence of the spin axial symmetry makes the applicability of the above explained spin
conservation analysis more sophisticated. A few particular cases are discussed below.

\subsection{Spin texturing in momentum space
           \label{ST-energy}
           }

If the TRS-invariant QSH system possesses the homogeneous Rashba SOI in the bulk, the spins of the
edge electron remain antiparallel only for two states of the Kramers doublet of the same energy
(i.e. for two modes with momentums $ \pm k $)
while the direction of the spin quantization depends on the electron energy
(i.e. on $ | k | $), the so-called spin-texturing
(ST) \cite{rod_2015}. In that case and unlike the Hamiltonian Eq.(\ref{H0}),
one should distinguish the chiral and spin bases of the fermions. These bases are related by the unitary
transformation which respects TRS: $ \tilde{\Psi} = \{ \psi_R, \psi_L \}^{\rm T}
= \hat{B}^\dagger_k \hat{\Psi}, \ \hat{B}^{\dagger}_k \hat{B}_k~=~1, \ \hat{B}_k = \hat{B}_{-k} $.
The chiral basis is needed to calculate the charge current carried by the helical electrons.
The Hamiltonian, which accounts for the ST in the momentum space, reads as:
\be
  \label{Hk_ST}
  H^{(k)}_{\rm ST} = v_F \sum_k k \, \hat{\Psi}^\dagger(k)
                                    \hat{B}_k \hat{\s}_z \hat{B}^{\dagger}_k \hat{\Psi}(k) .
\ee
As opposed to $ \hat{H}_0 $, the Hamiltonian $ H^{(k)}_{\rm ST} $ does not conserve z-component of the
electron spin. Therefore, in a general case, one cannot use the spin conservation analysis to
check the ballistic nature of the conductance. The obvious exception is the case of zero temperature where
transport is carried by the electrons at $ k_F $ and the product $ \hat{B}_{k_F} \hat{\s}_z \hat{B}^{\dagger}_{k_F} $
determines the direction of the spin quantization axis for the conduction electrons. Hence, the ST becomes
irrelevant at $ T = 0 $ because of the absence of the allowed phase space for the electrons and the nature
of the helical conductance (ballistic or subballistic) is fully determined by the properties of the SR
similar to the above discussed cases with the spin axial symmetry.

If temperature is finite, the ST violates the electron spin conservation and the helical conductance can become
subballistic. The well-known effect of the ST is the suppression of the helical conductance caused by the
inelastic (accompanied by the energy transfer) backscattering of the (disordered) helical electrons which
either interact with each other \cite{schmidt_2012,Mirlin-HLL,mcginley__2021} or experience an influence
of a noise \cite{vayrynen_noise-2018}. Less expected effect is an influence of the unscreened isotropic MI on
helical transport of the noninteracting electrons, see Ref.\cite{OYeVYu_2021} where the example of the spin-1/2
MI has been studied at $ T > T_K $. The underlying mechanism is the successive backscattering of the
helical electrons which have the same chirality but different energies. We remind the readers that, if there
is no ST, the unscreened isotropic or even U(1) symmetric spin-1/2 MI can backscatter one after another only
electrons of different chirality, thus, the dc conductance remains unaffected \cite{FurusakiMatveev}.

\subsection{Spin texturing in real space}

The spin texturing in real space originates in models where the direction
of the spin quantization on the helical edge is space-dependent. This could
result from the inhomogeneously changed direction of the SOI governed by a
combined effect of the spin-diagonal (Dresselhaus) and the spin-off-diagonal
(Rashba) SOIs \cite{rothe_2010,geissler_2014}. The latter does not violate
the topologically non-trivial state of the QSH sample but can be strong
on its edges. 
To describe the QSH edge with the spatially fluctuating SOI, one needs
an extended version of the Hamiltonian $ \hat{H}_0 $, Eq.(\ref{H0}),
which reads as follows:
\bea
\label{Hx_ST}
       \hat{H}^{(x)}_{\rm ST} & = & - i \frac{v_F}{2} \int {\rm d} x \ \hat{\Psi}^{\dagger}(x)
                  \left\{ \left( \vec{n}(x) \cdot \vec{\s} \right), \partial_x \right\}_+ \hat{\Psi}(x) = \cr
                              & = & 
                              - i v_F \int {\rm d} x \ \hat{\Psi}^{\dagger}(x)
                                \left( \vec{n}(x) \cdot \vec{\s} \right) \partial_x \hat{\Psi}(x) - 
                                \\
                              &   &
                                - i \frac{v_F}{2} \int {\rm d} x \ \hat{\Psi}^{\dagger}(x)
                                \left( \bigl[ \partial_x \vec{n}(x) \bigr] \cdot \vec{\s} \right) \hat{\Psi}(x)
                              \, ;
                              \nonumber
\eea
see Refs.\cite{strom_2010,budich_2012,geissler_2014,Matt_2016}.
In Eq.(\ref{Hx_ST}), $ \vec{\s} = \{ \hat{\s}_x, \hat{\s}_y, \hat{\s}_z \} $
and the anticommutator $ \{ \ldots \}_+ $ ensures the Hermiticity of
$ \hat{H}^{(x)}_{\rm ST} $.
In the Landauer setup of Fig.\ref{BlaclBoxSetUp}, the vector $ \vec{n} $ depends on the coordinate inside the SR and
is fixed by the equality $ \vec{n} = \{ 0, 0, 1 \} $ at contacts between the SR and the helical wires and in both wires.
Clearly, the Hamiltonian Eq.(\ref{Hx_ST}) coincides with $ \hat{H}_0 $, Eq.(\ref{H0}), if $ \vec{n} = \{ 0, 0, 1 \} 
= {\rm const} $ in the entire sample.
We emphasize that Eq.(\ref{Hk_ST}) and
Eq.(\ref{Hx_ST}) describe two principally different physical mechanisms of the ST and there is no simple
relation between these equations.
Using Eq.(\ref{Hx_ST}) in the QSH systems with the broken axial symmetry implies that
the product $ \hat{B}_{k_F} \hat{\s}_z \hat{B}^{\dagger}_{k_F} $ yields
the mean direction of the electron spin quantization axis and the SOI fluctuations
overwhelm the temperature-dependent effects of the ST at low temperatures,
$ T / {\cal E}_{st} \ll 1 $ ($ {\cal E}_{st} $ is the energy scale of the ST).

The electron spin density does not commute with Hamiltonian (\ref{Hx_ST}). Hence,
the total spin is not conserved in the laboratory frame and
one could presume that transport could be sub-ballistic even on clean
(without magnetic scatterers) edges with space-fluctuating SOI \cite{geissler_2014}.
This guess is incorrect. The role of the spin conservation, which is a kind of
masked in the model (\ref{Hx_ST}), for the protection of transport was noticed in
Ref.\cite{Matt_2016}. We will explicitly demonstrate this
in the next Section by using a simple and straightforward rotation of the fermionic basis.
We will use the spin conservation analysis in the rotated frame to make conclusive predictions
on the combined effects of the fluctuating SOI, the KHA and the electron-electron interactions.

\subsection{Combined influence of fluctuating SOI, KHA, \\
            and electron interaction on helical conductance \\
            - a case study
            \label{FluctSOI}
           }

Let us now explain how to uncover the spin conservation in the QSH edges with the
space-fluctuating (but static) direction of the spin quantization in the most straightforward way.
First, we explain the method by using the example of the clean Hamiltonian (\ref{Hx_ST}) and
further use the spin conservation analysis to study combined effects of the fluctuating
SOI and the scattering by the KHA in the presence of the electron-electron interaction.

The full Hamiltonian consists of five parts: $ \hat{H}_{\rm full} = \hat{H}^{(x)}_{\rm ST}
+ \hat{H}_{\rm e-KA} + \hat{{\cal H}}_H + \hat{H}_{\rm e-e} + \hat{H}_V $, where $ \hat{H}_{\rm e-KA} $
and $ \hat{H}_{\rm e-e} $ describe the exchange interaction of the helical electrons
with the KA and the electron-electron interaction, respectively, and are specified below
in Eqs.(\ref{H_e_KA},\ref{H_e-e}). $ \hat{{\cal H}}_H $ is the Hamiltonian of the Heisenberg
interaction between the MIs. In the case of the ZZ-interaction, $ \hat{{\cal H}}_H $ reduces
to Eq.(\ref{Hh}). In this section, we will explicitly take into account the in-plane
Heisenberg interaction, see Eq.(\ref{e-KHA_lab}) below.

For simplicity, we fix the strength of the SOI, such that $ | \vec{n}(x) | = 1 $,
and parameterize the coordinate-dependent direction of the SOI inside the SC by polar, $ \theta $,
and azimuthal, $ \varphi $, angles: $ \vec{n} = \{ \cos(\varphi) \sin(\theta), \sin(\varphi)
\sin(\theta), \cos(\theta) \} $. The boundary condition $
\vec{n} = \{ 0, 0, 1 \} $ implies that $ \theta = 0 $ at junctions of the SR with the connecting
helical wires. We further assume that the coordinate dependence of $ \vec{n} $ inside the
SC is slow on the scale of the lattice spacing.

Let us, at first, diagonalize the Hamiltonian (\ref{Hx_ST}) by using the
unitary transformation of the fermionic basis $ \tilde{\Psi} =
\hat{g}^\dagger \hat{\Psi} $ with the matrix $ \hat{g} \in \rm{SU(2)} $
being parameterized by the angles $ \{ \phi, \theta \} $:
\be
\label{g-matr}
    \hat{g} = e^{- i \frac{\varphi}{2} \hat{\s}_3} e^{- i \frac{\theta}{2} \hat{\s}_2}, \quad
    \hat{g}^\dagger \left( \vec{n} \cdot \vec{\s} \right) \hat{g} = \hat{\s}_3 .
\ee
$ \hat{g} = 1 $ outside the SR because of the boundary conditions for $ \vec{n} $.
Changing to the $ \tilde{\Psi} $-basis, we find
\be
 \label{Hrot-1}
       \hat{H}^{(x)}_{\rm ST}
                               = - v_F \int {\rm d} x \tilde{\Psi}^\dagger
          \left\{ i \hat{\s}_3 \p_x + \frac{1}{2} \cos(\theta) [\p_x \varphi] \right\} \tilde{\Psi} \, ;
\ee
see algebraic details in \appendixname~B.

The new frame, where the Hamiltonian (\ref{Hrot-1}) is defined, rotates inside the SR
together with the SOI. $ S_z $ spin component of the fermions is manifestly conserved
in this rotating frame. One can give the following interpretation
to that result: The electron with a given z-component of the spin enters
the SR where the electron spin follows the SOI direction. The initial direction
of the electron spin is restored after the electron leaves the SR and, thus,
the spin conservation law holds true because of the boundary conditions at the
contacts of the SR with the left/right helical wires.
The spin conservation in the rotating frame suggests that the helical conductance can remain ballistic
in spite of the absence of the global spin quantization axis in the laboratory frame. Indeed,
both terms in Eq.(\ref{Hrot-1}) are diagonal, the term containing $ \cos(\theta) $
is an effective forward-scattering potential; hence,
there is no backscattering in Eq.(\ref{Hrot-1}) and the Landauer conductance must
be ballistic, as expected \cite{Matt_2016}.

The protection of the helical conductance against the influence of the fluctuating SOI
on the clean and non-interacting edge is in general subtle and can be destroyed, e.g.,
by non-locality of the SOI fluctuations or by a curvature in the dispersion relation of
the helical electrons
\cite{kharitonov_2017}.
On a formal level, these phenomena break the spin conservation in the rotating frame.
Besides, the simple form of the diagonalized Hamiltonian (\ref{Hrot-1}) can be obtained
only for the stationary SOI fluctuations. If these fluctuations are dynamical,
the rotation by the matrix $ \hat{g} $ does not yield a conclusive information and
does not help to prove the ballistic nature of the helical conductance. For instance,
the model similar to Eq.(\ref{Hx_ST}) can appear due to the deformation
of the QSH edge by acoustic phonons \cite{groenendijk_2018}. In this case, the deviation
of the spin-quantization axis from z-direction is a dynamical degree of freedom, which
can be expressed via phonon operators. The theory of the electron-phonon coupling becomes
effectively non-local and the helical conductance becomes sub-ballistic at finite
temperatures due to the inelastic backscattering of the electrons by phonons \cite{groenendijk_2018}.

If the SOI fluctuations are static, the above described simple
though helpful method allows us to consider the SRs with- and without the
fermion scattering on equal footing. Let us gradually make the problem more complicated
and, at the next step, include into consideration the scattering caused by the KA. The
spin U(1)-invariant Hamiltonian of the exchange interaction between the itinerant electrons
and the MIs reads:
\bea
\label{H_e_KA}
 \hat{H}_{\rm e-KA} & = & \sum_j \hat{\Psi}^\dagger (x_j) (\vec S(j) \hat{J}
                   \cdot \vec{\sigma}) \hat{\Psi} (x_j) , \\
 \vec{S}(j) & = & \{ \tilde{S}_x(j), \tilde{S}_y(j), \hat{S}_z(j) \} \, .
\nonumber
\eea
We have introduced a diagonal matrix of the Kondo coupling constants
$ \hat{J} \equiv {\rm diag} ( J_K,  J_K, J_K^{(z)}) $. X- and Y-spin
components are related to the above introduced rotated spin operator
$ \tilde{S}_{x} = ( \hat{S}^+ +  \hat{S}^- ) / 2 $,
$ \tilde{S}_{y} = -i ( \hat{S}^+ -  \hat{S}^- ) / 2 $. Combined
effects of the randomly fluctuating SOI and a single MI has
been considered in Ref.\cite{kimme_2016}. The authors of this
paper have focused on the interplay of the Rashba disorder and
and off-diagonal anisotropy reflected by off-diagonal entries of
the matrix  $ \hat{J} $. We will show below that even a simpler
anisotropy of the XXZ-type (the diagonal matrix $ \hat{J} $ and the
U(1)-symmetric coupling in the laboratory frame) suffices to suppress
the helical conductance if the SOI fluctuates.

We re-write $ \hat{H}_{\rm e-KA} $ in the rotated fermionic
basis and simultaneously do an arbitrary unitary transformation of the MI
spin operators, $ \vec{S} \rightarrow {\cal S} $. $ \hat{H}_{\rm e-KA} $ in
the rotated bases reads as:
\be
\label{Hk-1}
   \hat{H}_{\rm e-KA} = \sum_j \tilde{\Psi}^\dagger (x_j) ({\cal S}(j) \tilde{J}
                   \cdot \vec{\sigma}) \tilde{\Psi} (x_j), \
                   \tilde{J} \equiv \hat{R}^T_{\cal S} \hat{J} \hat{R}_\Psi.
\ee
The orthogonal matrices $ \hat{R}_{{\cal S}, \Psi} \in \mbox{SO(3)} $ result from the unitary rotation
of the spins and the fermions, respectively.  $ \hat{R}_{\Psi} $ is parameterized by the
angles $ \{ \phi, \theta \} $ and $ \hat{R}_{{\cal S}} $ -- by another independent set of the
Euler angles, see \appendixname~C. Clearly, the z-component of the total (of the helical electrons
and of the MIs) spin is conserved in the rotated bases if (i) $ \tilde{J}_{kl}
= 0 $ for the entries $ \{k,l\} = \{1,3\}, \{2,3\}, \{3,1\}, \{3,2\} $;
(ii) $ \tilde{J}_{11} = \tilde{J}_{22} $; and (iii) $ \tilde{J}_{12} = - \tilde{J}_{21} $.
This is possible IFF $ J_K = J_K^{(z)} $, i.e. IFF the bare exchange interaction is isotropic.
After choosing $ \hat{R}_{\cal S} = \hat{R}_\Psi $, we reduce Eq.(\ref{Hk-1}) to the isotropic
version of the exchange interaction considered in Sect.\ref{e-KHA} at $ J_H = 0 $. Based on the
above discussion, we conclude that the conductance remains ballistic in the system where the
direction of the SOI rotates in the SR even if the SR contains the MIs isotropically
coupled (in the laboratory and rotating frames) to the conduction electrons.
Any anisotropy of the exchange coupling between the helical electrons and MIs
results in the violation of the spin conservation in the rotating frame and, thus,
removes the protection of the helical conductance.
For example, if the rotated coupling matrix has two coinciding finite diagonal elements $ \tilde{J}_{11}
= \tilde{J}_{22} \ne 0 $ but one non-zero off-diagonal element $ \tilde{J}_{13} \ne 0 $, the subballistic
correction to the helical conductance in the absence of the Kondo screening is $ \delta G_{\rm aniso} / G_0
\propto \tilde{J}_{11}^2 \tilde{J}_{13}^2 / ( \tilde{J}_{11}^2 + \tilde{J}_{33}^2 / 2 + \tilde{J}_{13}^2 ) $,
see the derivation and discussion in Ref.\cite{kurilovich_2017}.

The above consideration can be straightforwardly extended to the scattering of the
helical electron by the KHA. We start from the spin-preserving Heisenberg interaction
in the laboratory frame:
\bea
  \label{e-KHA_lab}
  \hat{{\cal H}}_H & = & \sum_j \left( \vec{S}(j) \hat{J}_H(j) \cdot \vec{S}(j+1) \right) \, , \\
  \hat{J}_H(j) & = & {\rm diag} ( J_H^\perp, J_H^\perp, J_H )_{x=x_j} \, .
  \nonumber
\eea
We rotate the spin degrees of freedom
\be
  \label{e-KHA_rot}
  \hat{{\cal H}}_H = \sum_j \left( {\cal S}(j) \hat{R}^T_S(j) \hat{J}_H(j) \hat{R}_S(j+1) \cdot {\cal S}(j+1) \right) \, .
\ee
If SOI variations are so smooth that the difference between the SOI at the space points
$ x_j $ and $ x_{j+1} $ can be neglected, we obtain $ \hat{R}^T_S(j) \hat{J}_H(j) \hat{R}_S(j+1) \simeq
\hat{R}^T_S(j) \hat{J}_H(j) \hat{R}_S(j) $ which reduces to the spin preserving form IFF
$ J_H^\perp(x_j) = J_H(x_j) $, cf. analysis of Eq.(\ref{Hk-1}). Thus, the total spin is
conserved in the rotating frame and, based on the results of Sect.\ref{e-KHA}, we conclude that
the ballistic helical conductance remains protected if both the Kondo- and Heisenberg couplings
are isotropic at each space point. If either the SOI variations are not smooth or there is a long range
Heisenberg interaction, the difference of the SOI at the points $ x_j $ and $ x_{j+1} $
(or $ x_j $ and $ x_{j+k} $ with $ k > 1 $ in the case of the long range MI interaction)
is not negligible, Eq.(\ref{e-KHA_rot}) violates the conservation of the total spin and the
helical conductance can be suppressed regardless of the isotropy of the coupling constants.

Including the density-density short-range electron interaction in this consideration
is also straightforward. In the case, the Hamiltonian $ \hat{H}_{\rm e-e}$ takes the
following form:
\be
  \label{H_e-e}
  \hat{H}_{\rm e-e} = \int {\rm d} x
          \left[
            V (\hat{\rho}_\uparrow + \hat{\rho}_\downarrow )^2 +
            \delta V \hat{\rho}_\uparrow \hat{\rho}_\downarrow
          \right]; \
   \hat{\rho}_{\uparrow,\downarrow} \equiv \psi_{\uparrow,\downarrow}^\dagger \psi_{\uparrow,\downarrow} .
\ee
We have singled out terms which are invariant, $ \sim V $ with $ V = {\rm const} $, and not-invariant ,
$ \sim \delta V $ with $ \delta V = {\rm const} $, under the unitary rotation by the matrix $ \hat{g} $. The
fermionic density operators in the rotated basis read as:
\bea
   \hat{\rho}_\uparrow & = &   \hat{\rho}_R \cos^2 \left( \frac{\theta}{2} \right)
                             + \hat{\rho}_L \sin^2 \left( \frac{\theta}{2} \right) - \\
                       &   & - \frac{ \sin(\theta) }{2} \left( \psi_R^\dagger \psi_L + \psi_L^\dagger \psi_R \right); \cr
   \hat{\rho}_\downarrow & = & \hat{\rho}_R \sin^2 \left( \frac{\theta}{2} \right)
                             + \hat{\rho}_L \cos^2 \left( \frac{\theta}{2} \right) + \\
                       &   & + \frac{ \sin(\theta) }{2} \left( \psi_R^\dagger \psi_L + \psi_L^\dagger \psi_R \right);
                       \nonumber
\eea
here $ \hat{\rho}_{R/L} = \psi_{R/L}^\dagger \psi_{R/L} $ with $ \psi_{R/L} $ being entries of the spinor $ \tilde{\Psi} $.
The invariant part of $ \hat{H}_{\rm e-e} $ retains its form in the rotated basis:
\be
   \hat{H}_{\rm inv} = V \int {\rm d} x \, (\hat{\rho}_R + \hat{\rho}_L )^2 .
\ee
$ \hat{H}_{\rm inv} $ does not break the spin conservation in the rotated basis and cannot lead to the
suppression of the helical conductance. The non-invariant part of $ \hat{H}_{\rm e-e} $ generates after
the basis transformation all possible types of the electron interaction in 1D systems \cite{Giamarchi},
including nontrivial umklapp processes in the SR
\footnote{We have used the conventional order of space arguments in all parts of Eq.(\ref{umklapp}), for
example $ \psi_R^\dagger(x) \psi_R^\dagger(x+a_0) \psi_L(x+a_0) \psi_L(x) $, where
the shift of two arguments by the lattice constant $ a_0 $ is the standard regularization}:
\bea
   \label{umklapp}
   \hat{H}_{\rm um} & = & \frac{\delta V}{4} \int {\rm d} x
          \left[
             - \psi_R^\dagger \psi_R^\dagger \psi_L \psi_L \sin^2(\theta) +
          \right. \\
                    & + &
          \left.
              \left(
                \psi_R^\dagger \psi_L^\dagger \psi_R \psi_R -
                \psi_L^\dagger \psi_R^\dagger \psi_L \psi_L
              \right) \sin(2 \theta) + h.c.
          \right] .
          \nonumber
\eea
This is similar to the generation of the effective umklapp due to the ST in the momentum space,
cf. Ref.\cite{Mirlin-HLL}. Since umklapp (two-particle) scattering does not preserve the total
spin of the helical fermions
it can suppress the helical conductance \cite{WuBernevigZhang,Mirlin-HLL,Oreg-MPBS}.

We would like to note that
using the spin conservation analysis presented in this Section helps to avoid the excessive complexity
inherent in the heavy theoretical machinery and can prevent erroneous conclusions on the suppression
of ballistic helical transport in the spin preserving setups, cf. Ref.\cite{geissler_2014,strom_2010}.

\section*{Summary}

We have presented a general approach for the analysis of helical
transport in Quantum Spin Hall systems. Based on this approach,
we have critically reviewed and discussed theories addressing
protection of this transport and studied in detail particular
examples.

Though the existence of
helical modes on edges of QSH samples is protected by the nontrivial
topology of the bulk and the time-reversal symmetry, the edge
electrons are vulnerable to backscattering processes caused by
various interactions. However, the backscattering does not necessarily
suppress the ballistic helical conductance. For instance, the helical
conductance is ballistic despite backscattering in the QSH systems with
the axial electron spin symmetry if the projection of the total
spin on this symmetry axis is conserved. We have extended the
discussion of the general approach by a detailed study of the
new example where the helical conductance remains ballistic in the spin
preserving setup in spite of the scattering of the itinerant electrons
by the array of Heisenberg-interacting magnetic impurities. Interestingly,
a magnetic spin ordering of the Ising type, which formally breaks
the time reversal symmetry on long time intervals, also has no
influence on the helical conductance if the total spin is conserved.

If the spin axial symmetry is broken and the so called spin texturing
arises on the QSH edges, the spin conservation analysis is more sophisticated
and not always applicable. One should distinguish here two physically
different cases. If the broken spin axial symmetry results from an
(effective) Rashba spin-orbit interaction in the bulk, one comes across the
spin texturing in the momentum space which breaks the spin conservation
law though is itself unable either to backscatter or to affect the helical
conductance. Being combined with various sources
of the backscattering, such a ST is able to suppress the helical
conductance.
These effects of the broken
axial symmetry are often frozen out in the experimentally relevant range
of low temperatures and, therefore, cannot dominate suppression of the
helical conductance.

Another class of models with the broken axial symmetry is given by the
systems where the spin-orbit interaction strongly fluctuates on the QSH
edge. This leads to the spin texturing in the real space, i.e., to
spatial fluctuations of the spin quantization axis. The spin conservation
law is violated in the laboratory frame by the SOI fluctuations, however,
the spin conservation analysis is applicable in the frame which rotates
in space together with the SOI direction. Using this approach, we have
considered previous unexplored combined effects of the ST in the real
space and scattering of the (non-interacting and interacting) helical
electron by the Kondo-Heisenberg array of the MIs. We have identified
conditions under which the helical conductance in these systems is protected
even at finite temperatures.

To conclude, we have demonstrated that the spin conservation analysis
is a useful tool for the study of helical transport in a variety
of the QSH systems. The setups studied in this paper in details show
that this approach helps
to identify physical mechanisms being relevant or irrelevant for the
suppression of the helical conductance.

\begin{acknowledgments}
O.M.Ye. acknowledges support from the DFG through the grant YE 157/2-3.
V.I.Yu. acknowledges support from the Basic research program of HSE.
\end{acknowledgments}

\bibliography{Bibliography,TI,Combined-SOI}

%
%
%
%

\widetext

\newpage

\appendix

\section{Helical wire coupled to a Kondo-Heisenberg array \label{Ideal_Cond-App}}

Non-local Matsubara conductivity of a helical wire can be expressed in terms of the Green's function (GF) of
bosonized excitations \cite{Giamarchi}:
\be
\label{NonLocSigma}
   \sigma(x,x'; \bar{\omega} ) = ( e^2 \bar{\omega} / \pi^2 ) \, G(x,x'; \bar{\omega} ) \, .
\ee
We need the generating functional, $ Z[\chi] $, which allows one to calculate $ G(x,x'; \bar{\omega} ) $
for the helical wire coupled to localized spins (a Kondo-Heisenberg array). Consider first {\it the spin
preserving setup}, where this generating functional reads as follows:
\begin{eqnarray}
\label{GenFunc}
    Z[\chi] & = & \frac{1}{Z[0]} \int {\cal D}\{ n_z , \tilde{\alpha} \}
                         \exp\Bigl( - S_{\rm MI}[ n_z, \tilde{\alpha} ] \Bigr) \,
                   \int {\cal D}\{ \phi \}
                      \exp\Bigl( - S_{\rm HLL}[\phi, \chi] - S_{\rm WZ} [ n_z, 2 \phi] \Bigr) \, ; \\
\label{VarDeriv}
    G(x_1,x_2; \tau_1 - \tau_2 ) & = & \frac{\delta^2 Z[\chi] }{ \delta \chi(\zeta_1) \delta \chi(\zeta_2) } \Bigl|_{\chi \to 0 } \, , \quad
    \zeta_{1,2} \equiv \{ x_{1,2}, \tau_{1,2} \} \, .
\end{eqnarray}
Here $ Z[0] $ is the partition function; actions $ S_{{\rm HLL}, {\rm WZ}, {\rm MI}} $ correspond to
the Lagrangians $ {\cal L}_{{\rm HLL}, {\rm WZ}, {\rm MI}} $ (see Sect.\ref{e-KHA}, Eqs.(\ref{L_HLL}--\ref{Lagr})):
\bea
   S_{\rm HLL} & = & \int {\rm d} \zeta \left\{
                \frac{1}{2\pi v_F} \left[ (\p_\tau \phi)^2 + (v_F \p_x \phi)^2 \right] +
                i \chi \phi
                                                    \right\}  \, ; \\
   S_{\rm WZ} [ n_z, 2 \phi] & = & 2 i s \int {\rm d}  \zeta \, \rho_s (1 - n_z ) \partial_\tau \phi =
                                                2 i s \int {\rm d}  \zeta \, \rho_s \phi \, \partial_\tau n_z \, ; \\
\label{Sks}
   S_{\rm MI} & = & \int {\rm d}  \zeta \, \rho_s(x)
          \left( 2 s  J_{K}(x) \sqrt{1 - n^2_z} \cos( \tilde{\alpha} )
                 + s^2 J_H(x) n_z(x,\tau) n_z(x+\xi_s,\tau)
          \right)
                       + S_{\rm WZ} [ n_z, \tilde{\alpha} ] \, .
\eea

The Gaussian integral over $ \phi $ in Eq.(\ref{GenFunc}) can be calculated straightforwardly:
\bea
 \label{EffAct}
 \frac{ \displaystyle \int {\cal D}\{ \phi \} \exp\Bigl( - ( S_{\rm HLL}[\phi, \chi] + S_{\rm WZ} [ n_z, 2 \phi] ) \Bigr) }
        { \displaystyle \int {\cal D}\{ \phi \} \exp\Bigl( - ( S_{\rm HLL}[\phi, \chi = 0 ]  ) \Bigr) } & = &
                    \cr
        \exp \Biggl( - S_{\rm zz} + \frac{1}{2} \int {\rm d} \zeta {\rm d} \zeta' \,
                   \Bigl[ \chi(\zeta) G_0(\zeta-\zeta', \omega) \chi(\zeta') & + &
                             2 s \chi(\zeta) G_0(\zeta-\zeta') \rho_s(x') \p_{\tau'} n_z(\zeta')
                   \Bigr]
                \Biggr) \, ;
\eea
with
\be
\label{Szz-time}
       S_{\rm zz} \equiv - \int {\rm d} \zeta {\rm d} \zeta' \,
                   \Bigl[ 2 s^2 \rho_s(x) \p_{\tau} n_z(\zeta) G_0(\zeta-\zeta') \rho_s(x') \p_{\tau'} n_z(\zeta') \Bigr] \, .
\ee
The variational derivative over the source field, Eq.(\ref{VarDeriv}), yields:
\begin{eqnarray}
\label{GF-eq1}
   G(x_1,x_2; \tau_1 - \tau_2) = G_0( \zeta_1 - \zeta_2 ) +
                 \int {\rm d} \zeta {\rm d} \zeta' \, G_0( \zeta_1 - \zeta ) \rho_s(x)
                              \bigl[ \p^2_{\tau,\tau'} {\cal C}_{zz}(\zeta, \zeta') \bigr]
                              \rho_s(x') \, G_0( \zeta' - \zeta_2) \, ;
\end{eqnarray}
where
\begin{equation}
\label{N_z-z}
  {\cal C}_{zz}(\zeta, \zeta') \equiv \langle \langle \bigl[ s n_z(\zeta) \bigr]
                                                 \bigl[ s n_z(\zeta') \bigr]
                                      \rangle \rangle_{S_{\rm MI} + S_{\rm zz}} \, ;
\end{equation}
$ G_0 ( \zeta_1 - \zeta_2 ) $ is the bare bosonic GF of the clean wire, and $ \langle \langle {\cal A} {\cal B} \rangle \rangle \equiv
\langle {\cal A} {\cal B} \rangle - \langle {\cal A} \rangle \langle {\cal B} \rangle$. Decoupled part of $ {\cal
C}_{zz} $ does not contribute to Eq.(\ref{GF-eq1}) because $ \p_\tau \langle n_z \rangle = 0 $.
The averaging in Eq.(\ref{N_z-z}) is performed over the full spin action $ S_{\rm MI} + S_{\rm zz} $.
If there is no exchange interaction between the itinerant electrons and the localized spins,
$ J_K = 0 $, Eq.(\ref{GF-eq1}) manifestly reproduces $ G_0 ( \zeta_1 - \zeta_2 ) $ because the MI spins are not coupled and
$ \p^2_{\tau,\tau'} {\cal C}_{zz}(\zeta, \zeta') = 0 $.

Integrating by parts and using transnational invariance of $ G_0 ( \zeta_1 - \zeta_2 ) $, we find:
\be
\label{GF-eq2}
   G(x_1,x_2; \tau_1 - \tau_2) = G_0( \zeta_1 - \zeta_2 ) +
                \int {\rm d} \zeta {\rm d} \zeta' \, \bigl[ \p_{\tau_1} G_0( \zeta_1 - \zeta ) \bigr] \rho_s(x)
                {\cal C}_{zz}(\zeta, \zeta') \rho_s(x') \, \bigl[ \p_{\tau_2} G_0( \zeta' - \zeta_2) \bigr] \, .
\ee
The expression for $ G_0 $ in the momentum-frequency representation
reads:
\begin{equation}
   \label{G0_bos}
   G_0(q, {\bar \omega}) = - \langle \phi^*(q, {\bar \omega}) \phi(q, {\bar \omega}) \rangle =
                           - \frac{ \pi v_F }{ {\bar \omega}^2 + (v_F q)^2 } \, .
\end{equation}
We are interested in the dc response of the wire at zero temperature. Changing from the momentum
to the coordinate, we obtain in the low-frequency limit, $ |\bar{\omega} (x_1-x_2) | / v_F \ll 1 $,
the equality:
\begin{equation}
\label{GF0}
   \bar{\omega} \, G_0(x_1-x_2; \bar{\omega} ) = - \frac{\pi}{2} {\rm sign}(\bar{\omega}) \, .
\end{equation}
Let us now Fourier-transform Eq.(\ref{GF-eq2}) for the GF, analytically continue it to
the upper half-plane to obtain the physical retarded correlation function, $ G^R $, and simplify it
in the low frequency limit by using Eq.(\ref{GF0}):
\be
\label{GF-eq3}
   G^R(x_1,x_2; \omega ) = G_0^R ( x_1 - x_2; \omega ) + \delta G^R( \omega ) ;
\ee
where
\be
  \delta G^R(\omega) = \left( \frac{\pi}{2} \right)^2 {\cal S}^{\rm tot}_{zz}(\omega) ; \quad
     {\cal S}^{\rm tot}_{zz}(\omega) \equiv \int {\rm d} x_{1} {\rm d} x_{2} \, \rho_s(x_1) {\cal C}_{zz}^R(x_1, x_2; \omega ) \rho_s(x_2) \, .
\ee

We have to analyze the low frequency limit of the product $ \omega \delta G^R(\omega) $ which yields a correction to
the nonlocal conductivity of the clean wire, see Eq.(\ref{NonLocSigma}). We note that $ {\cal S}^{\rm tot}_{zz} \, $ is
the retarded correlation function of the KHA total spin. Since all MIs are located inside the finite SR, the total spin
of the KHA is also finite. We conclude that the retarded correlation $ {\cal S}^{\rm tot}_{zz}(\omega) $ is bounded
and, therefore, the product $ \omega \, {\cal S}^{\rm tot}_{zz}(\omega) $ vanishes in the dc limit. Hence, we arrive at
the trivial limit
\be
\label{Answer}
  \lim_{\omega \to 0 } \left( \omega \, \delta G^R(\omega) \right) = 0.
\ee
Eq.(\ref{Answer}) proves that the dc conductance of the helical wire coupled to {\it any spin conserving finite
Kondo-Ising array} (i.e. the KHA with only ZZ-coupling of the MI spins)
coincides with the ballistic conductance of the clean helical wire regardless of (i)~properties of contacts
between the wire and the region of localized spins, (ii)~a spatial inhomogeneity or a spin disorder of the KHA, etc.

If the spin array is infinite, the total spin of the KHA is not bounded, $ {\cal S}^{\rm tot}_{zz} $ may diverge, our
approach is not applicable any longer and the theory of Ref.\cite{AAY} for the infinite Kondo array should be used instead.
This theory predicts the dc conductivity (not conductance!) with the renormalized Drude weight.

The action in Eq.(\ref{GenFunc}) is drastically changed in the spin non-preserving setup. For example, if the coupling
of the MIs and the itinerant electrons breaks the spin U(1) symmetry, one cannot obtain the simple equations of the type
(\ref{EffAct},\ref{GF-eq1}). The action for the charge sector acquires the form of a Sine-Gordon
theory, where the charge density waves can be pinned. This leads to the suppression of the ballistic helical conductance,
cf. Refs.\cite{AAY,YeWYuA}.

\section{Rotation of spinors \label{RotatedSpinors}}

Consider the Hamiltonian Eq.(\ref{Hx_ST}):
\begin{eqnarray}\label{H}
\hat{H}^{(x)}_{\rm ST} = -\frac{i}{2}v_F\int dx \,
\hat{\Psi}^{\dagger}(x)\left\{( \vec{n} \cdot \vec{\sigma}), \partial_{x}\right\}_{+}\hat{\Psi}(x)
= -iv_F \int dx \,\hat{\Psi}^{\dagger}(x)
\left[
(\vec{n} \cdot \vec{\sigma})\partial_{x}
+ \frac{1}{2}[\partial_{x}(\vec{n} \cdot \vec{\sigma})]
\right]\hat{\Psi}(x)
 \, .
\end{eqnarray}
Here $\mathbf{n}= \left\{\sin{(\theta)}\cos{(\varphi)}, \sin{(\theta)}\sin{(\varphi)}, \cos{(\theta)}\right \}$ is the unit vector.
Let us introduce a unitary matrix
\begin{eqnarray}\label{g}
\hat{g} = \exp \Bigl[ -i\frac{\varphi}{2} \hat{\sigma}_3 \Bigr]
          \exp \Bigl[ -i\frac{\theta}{2}  \hat{\sigma}_2 \Bigr] \, ,
\end{eqnarray}
which has the property
\begin{eqnarray}\label{gng}
\hat{g}^{\dag}(\vec{n} \cdot \vec{\sigma})\hat{g} = \hat{\sigma}_3 \, \Leftrightarrow \,
\hat{g}\hat{\sigma}_3 \hat{g}^{\dag}=(\vec{n} \cdot \vec{\sigma}) .
\end{eqnarray}
Changing to a new spinor $\tilde{\Psi}=\hat{g}^{\dag}\hat{\Psi}$, we obtain
the new Hamiltonian in the form
\begin{eqnarray}\label{H-rot}
\hat{H}_{\mathrm{rot}} = -iv_F \int dx \,\tilde{\Psi}^{\dagger}(x)\hat{g}^{\dag}
\left[( \vec{n} \cdot \vec{\sigma})\partial_{x}
+ \frac{1}{2}[\partial_{x}( \vec{n} \cdot \vec{\sigma})]
\right]\hat{g}\tilde{\Psi}(x)
= -iv_F \int dx \,\tilde{\Psi}^{\dagger}(x)
\left[\hat{\sigma}_3\partial_{x} + \hat{h}
\right]\tilde{\Psi}(x)
 \, ,
\end{eqnarray}
where
\begin{eqnarray}\label{h}
\hat{h} = \hat{g}^{\dag}( \vec{n} \cdot \vec{\sigma} )\hat{g}
\hat{g}^{\dag}\partial_{x}\hat{g}
+\frac{1}{2}\hat{\sigma}_3 ( \partial_{x}\hat{g}^{\dag} ) \hat{g}
+ \frac{1}{2}\hat{g}^{\dag}\partial_{x}\hat{g}\hat{\sigma}_3
= \frac{1}{2}\left[
\hat{\sigma}_3 \hat{g}^{\dag}\partial_{x}\hat{g} +
\hat{g}^{\dag}\partial_{x}\hat{g}\hat{\sigma}_3
\right]
 \, .
\end{eqnarray}
We have used Eq.(\ref{gng}) to obtain Eq.(\ref{h}). The spatial derivative of the matrix
$\hat{g}(x)$ reads as
\begin{eqnarray}\label{gx}
\partial_{x}\hat{g} = -i\frac{1}{2}
\left[
 \hat{\sigma}_3\hat{g} \partial_{x}\varphi +
 \hat{g}\hat{\sigma}_2 \partial_{x}\theta
\right]
\, .
\end{eqnarray}
Inserting this expression in Eq.(\ref{h}), we find
\begin{eqnarray}\label{h1}
\hat{h} =
-\frac{i}{4}
\left[\hat{\sigma}_3\hat{g}^{\dag}\hat{\sigma}_3\hat{g}
+ \hat{g}^{\dag}\hat{g}\hat{\sigma}_3
\right] \partial_{x}\varphi
=
-\frac{i}{4}
\hat{g}^{\dag}\left[
\hat{g}\hat{\sigma}_3\hat{g}^{\dag}\hat{\sigma}_3
+ \hat{\sigma}_3\hat{g}\hat{\sigma}_3\hat{g}^{\dag}
\right]
\hat{g} \partial_{x}\varphi
\, .
\end{eqnarray}
Using Eq.(\ref{gng}), we simplify Eq.(\ref{h1}):
\begin{eqnarray}\label{h-final}
\hat{h} = -\frac{i}{4}[ (\vec{n} \cdot \vec{\sigma}) \hs_3 + \hs_3 (\vec{n} \cdot \vec{\sigma})] \partial_{x}\varphi =
-\frac{i}{2}\cos{(\theta)}\partial_{x}\varphi \, .
\end{eqnarray}
Thus, the Hamiltonian in the rotated frame takes the form
\begin{eqnarray}\label{H-rot-final}
\hat{H}_{\mathrm{rot}} =
- v_F \int dx \,\tilde{\Psi}^{\dagger}(x)
\left[ i\hat{\sigma}_3\partial_{x} + \frac{1}{2}\cos{(\theta)}\partial_{x}\varphi
\right]\tilde{\Psi}(x)
 \, ,
\end{eqnarray}
see Eq.(\ref{Hrot-1}) in the main text.

\section{Electron-MI interaction in the rotated basis and the spin conservation}

The unitary transformation of the electron spin operators can be re-written as
an expansion in Pauli matrices:
\be
\label{Psi-ident}
  \hat{g}^\dagger \vec{\s}^T \hat{g} = \hat{R}_\Psi \vec{\s}^T, \ \vec{\s} = \{ \hat{\s}_1, \hat{\s}_2, \hat{\s}_3 \} .
\ee
The matrix $ \hat{g} \in SU(2) $ is defined in Eq.(\ref{g-matr}) of the main text.
Entries of the matrix $ \hat{R}_\Psi \in SO(3) $ are coefficients of this expansion.

A similar expansion can be done for the unitary transformation of the MI spin operators:
\be
\label{S-ident}
  \hat{g}_{\cal S}^\dagger \vec{S} \hat{g}_{\cal S} = \vec{S} \hat{R}^T_{\cal S}, \ \vec{S} = \{ \tilde{S}_{x}, \tilde{S}_{y}, \hat{S}_{z} \} .
\ee
with matrices $ \hat{g}_{\cal S} \in SU(2) $, $ \hat{R}_{\cal S} \in SO(3) $. For simplicity,
we assume that the MIs also have spin 1/2. The matrix $ \hat{g}_{\cal S} $ can be parameterized
by three Euler angles:
\be
\label{g-s}
\hat{g}_{\cal S} = \exp \Bigl[ -i\frac{\varphi_S}{2} \hat{\sigma}_3 \Bigr]
            \exp \Bigl[ -i\frac{\theta_S}{2}  \hat{\sigma}_2 \Bigr]
            \exp \Bigl[ -i\frac{\psi_S}{2} \hat{\sigma}_3 \Bigr] \,
\ee
[the generalization to the arbitrary MI spin can be done by substituting $ \tilde{S}_y $
and $ \hat{S}_z $ for $ \hat{\s}_{2,3} / 2 $, respectively, in Eq.(\ref{g-s})].
Eqs.(\ref{Psi-ident}-\ref{S-ident}) have been used to obtain the expression for the effective
coupling constant in the main text, $ \, \tilde{J} \equiv \hat{R}^T_{\cal S} \ {\rm diag}
\left( J_K, J_K, J_K^{(z)} \right) \hat{R}_\Psi $, see Eq.(\ref{Hk-1}).

The z-component of the total spin is conserved if $ \hat{H}_{\rm e-KA} $, Eq.(\ref{Hk-1}), does not contain
terms $ \hat{S}_{z} \hat{\s}^{\pm} $, $ \hat{S}^{\pm} \hat{\s}_{3} $, $ \hat{S}^{+} \hat{\s}^{+} $, and
$ \hat{S}^{-} \hat{\s}^{-} $ in the rotated basis. This requires the following properties of $ \, \tilde{J} $:
\bea
\label{Req-i}
 \tilde{J}_{kl} & = & 0 \ \mbox{ for the entries } \ \{k,l\} = \{1,3\}, \{2,3\}, \{3,1\}, \{3,2\} ; \\
\label{Req-ii}
 \tilde{J}_{11} & = & \tilde{J}_{22} ; \\
\label{Req-iii}
 \tilde{J}_{12} & = & - \tilde{J}_{21} .
\eea
Eqs.(\ref{Req-i}-\ref{Req-iii}) result in a set of transcendental equations relating
the Euler angles $ \{ \varphi_S, \theta_S, \psi_S \} $ and $ \{ \varphi, \theta \} $
at given bare coupling constants $ J_K, J_K^{(z)} $. The spin can be conserved only
if all these requirements are compatible with each other. Let us check the compatibility
of equations governed by Eq.(\ref{Req-i}). After lengthy but straightforward
trigonometric manipulations, they can be reduced to the following form:

\bea
\label{J13}
   \tilde{J}_{13} & = &
      J_K \sin( \theta ) \left[ \cos( \theta_S ) \cos( \psi_S ) \cos( \phi - \phi_S ) + \sin( \psi_S ) \sin( \phi -\phi_S ) \right] -
      J_K^{(z)} \cos( \theta ) \sin( \theta_S ) \cos ( \psi_S ) = 0 ; \\
\label{J31}
   \tilde{J}_{31} & = &
      J_K \cos( \theta ) \sin( \theta_S ) \cos( \phi - \phi_S )- J_K^{(z)} \sin( \theta ) \cos( \theta_S ) = 0 ; \\
\label{J23}
   \tilde{J}_{23} & = & J_K \sin( \theta ) \left[ \cos( \psi_S ) \sin( \phi - \phi_S ) - \cos( \theta_S ) \sin( \psi_S ) \cos( \phi - \phi_S ) \right] +
                        J_K^{(z)} \cos( \theta ) \sin( \theta_S ) \sin( \psi_S ) = 0 ; \\
\label{J32}
   \tilde{J}_{32} & = & J_K \sin( \theta_S ) \sin( \phi_S - \phi ) = 0 .
\eea
Eq.(\ref{J32}) implies that either $ \theta_S = 0 $ or $ \phi_S = \phi $. The former equality is incompatible
with Eq.(\ref{J31}) (except for the trivial case $ \theta = 0 $ which corresponds to the fixed direction of SOI along
the z-axis). Thus, the total spin can be conserved in the rotated basis IFF $ \phi_S = \phi $. In this case,
Eqs.(\ref{J13}-\ref{J23}) reduce to the following form:
\bea
\label{J13-1}
   \tilde{J}_{13} & = &
      \left[ J_K \sin( \theta ) \cos( \theta_S ) - J_K^{(z)} \cos( \theta ) \sin( \theta_S ) \right] \cos ( \psi_S ) = 0 ; \\
\label{J31-1}
   \tilde{J}_{31} & = &
      J_K \cos( \theta ) \sin( \theta_S ) - J_K^{(z)} \sin( \theta ) \cos( \theta_S ) = 0 ; \\
\label{J23-1}
   \tilde{J}_{23} & = & \left[ - J_K \sin( \theta ) \cos( \theta_S ) + J_K^{(z)} \cos( \theta ) \sin( \theta_S ) \right] \sin( \psi_S ) = 0 .
\eea
Eqs.(\ref{J13-1}) and (\ref{J23-1}) are compatible IFF the expression in square brackets is equal to zero.
Thus the spin conservation is possible IFF
\bea
  J_K \sin( \theta ) \cos( \theta_S ) & = & J_K^{(z)} \cos( \theta ) \sin( \theta_S ) \, ; \\
  J_K \cos( \theta ) \sin( \theta_S ) & = & J_K^{(z)} \sin( \theta ) \cos( \theta_S ) \, .
\eea
These equations are compatible IFF $ |J_K| = J_K^{(z)} $. The sign of $ J_K $ is irrelevant
(this follows from a possibility to change signs of $ \tilde{S}_{x,y} $ and $ J_K $ simultaneously).
Therefore, the isotropic bare coupling of the itinerant electrons and the MIs, $ J_K = J_K^{(z)} $,
is the necessary condition for the spin conservation in the example considered in Sect.\ref{FluctSOI}.

%
%

\end{document}